\documentclass[fleqn,10pt,table]{article} 
\usepackage[utf8]{inputenc}
\usepackage[english]{babel} 
\usepackage[style=numeric,sorting=nyt,backend=biber]{biblatex}
\usepackage{csquotes}
\usepackage{diagbox}
\usepackage[]{graphicx, xcolor}
\usepackage{microtype}
\usepackage{booktabs}
\usepackage{hyperref}
\usepackage{amssymb}
\usepackage{amsmath}

\usepackage{hyperref} 
\hypersetup{breaklinks=true,pdftitle={Assessing the Auditability of AI-integrating Systems},pdfauthor={Linda Fernsel; Yannick Kalff; Katharina Simbeck}}

\title{Assessing the Auditability of AI-integrating Systems: A Framework and Learning Analytics Case Study} 

\author{Linda Fernsel\textsuperscript{1}, Yannick Kalff\textsuperscript{1}, Katharina Simbeck\textsuperscript{1}\\\textsuperscript{1} HTW University of Applied Sciences, Berlin, Germany\\\href{mailto:yannick.kalff@htw-berlin.de}{yannick.kalff@htw-berlin.de}, \href{mailto:simbeck@htw-berlin.de}{simbeck@htw-berlin.de}} 
\begin{document}


\maketitle 





\section{Abstract}
Audits contribute to the trustworthiness of Learning Analytics (LA) systems that integrate Artificial Intelligence (AI) and may be legally required in the future. We argue that the efficacy of an audit depends on the auditability of the audited system. Therefore, systems need to be designed with auditability in mind. We present a framework for assessing the auditability of AI-integrating systems that consists of three parts: (1) Verifiable claims about the validity, utility and ethics of the system, (2) Evidence on subjects (data, models or the system) in different types (documentation, raw sources and logs) to back or refute claims, (3) Evidence must be accessible to auditors via technical means (APIs, monitoring tools, explainable AI, etc.). 
We apply the framework to assess the auditability of Moodle’s dropout prediction system and a prototype AI-based LA. We find that Moodle's auditability is limited by incomplete documentation, insufficient monitoring capabilities and a lack of available test data.
The framework supports assessing the auditability of AI-based LA systems in use and improves the design of auditable systems and thus of audits.

\textbf{Keywords}: audit, auditability, artificial intelligence, learning analytics

\section{Introduction} 
\label{sec:1}

Artificial Intelligence (AI) significantly impacts the field of Learning Analytics (LA). LA itself is gaining relevance in higher education \cite{Baek.2023}, K-12 classes \cite{Paolucci.2024}, and for virtual education \cite{Elmoazen.2023}. AI adds to the utility of LA elements of educational data mining \cite{Baek.2023}, deep learning capabilities, machine learning \cite{Ouyang.2023}, or predictive and prescriptive analytics \cite{Sghir.2022, Xiong.2024}. AI in LA offers data-driven insights into learning processes and students' behavior to predict learning success, risks of failure or drop-out, and to prescribe proactive measures \cite{Alfredo.2024, Susnjak.2024}. Above that, AI technologies promise opportunities to improve learning situations and outcomes especially for disadvantaged and struggling students \cite{Khalil.2023}. 

However, AI in LA entail ethical issues \cite{Rzepka.2022, Rzepka.2023} and their utility or maturity in education are unclear \cite{Drugova.2024}. The potential threat to equality and equity principles in education raised research and practitioner interest to mitigate discriminatory effects of AI models in LA \cite{Simbeck.2023, Rzepka.2023}. Above that, ethical concerns create the urgency for adequate legislation to counter negative effects or prevent biased systems. AI and AI-driven LA face regulation, such as the European AI Act \cite{EuropeanUnion.2024}, or frameworks to impose ethical requirements on AI products \cite{Toreini.2022, Slade.2019}, and to mitigate potential negative effects and discriminatory biases \cite{Baker.2021, Prinsloo.2017}. For this case, the “AI Act” \cite{EuropeanUnion.2024} aims to regulate “high-risk” AI systems, like AI-based LA because of their impact on personal educational success. 

Regulatory frameworks, including the AI Act, require audits of AI systems that certify their legal and ethical compliance \cite{Berghoff.2022, Toreini.2022}. The AI Act mandates two types of audits for high-risk AI systems: conformity assessments before deployment and post-market monitoring after system deployment \cite{EuropeanUnion.2024}. Audits provide accountability and transparency, which is also necessary for establishing trust in AI systems \cite{Williams.2022, Bose.2019}. On a practical level, audits allow stakeholders, such as system providers or deploying institutions, regulators, and subjects of the systems’ decisions, to understand how the system decides and to identify and correct biases or errors \cite{Springer.2019}. Audits of AI-based LA systems are effective in discovering and tackling algorithmic bias \cite{Rzepka.2023}. However, audits struggle with systems that are inaccessible, opaque, or proprietary \cite{Fernsel.2024}. We argue that AI-based LA systems must be auditable for any audit to achieve its results. Therefore, our research questions are:
\begin{description}
\item[RQ1] How does a framework to assess the auditability of AI-based LA support auditors to judge a given system?
\item[RQ2] How does a framework to assess auditability improve the development of AI-based LA?
\end{description}

We define AI systems as software systems that implement methods of machine learning. AI-based LA systems implement machine learning methods to leverage learning data for analysis, predictions, and prescriptions in educational contexts \cite{Baek.2023, Ouyang.2023}.

In section~\ref{sec:2}, we define \textit{auditability} and discuss the limitations of auditability of AI-integrating LA systems. Auditability requires a system to collect evidence about specific claims and make it accessible for independent assessments \cite{Weigand.2013, Wolnizer.2006}. In section~\ref{sec:3}, we present our framework for assessing the auditability of AI-based LA systems. We apply our framework to Moodle’s AI-based LA system (section~\ref{sec:4}) and a prototype LA (section~\ref{sec:5}). In section~\ref{sec:6}, we discuss the implications and limitations of our work and conclude in section~\ref{sec:7}.

\section{Literature Review}
\label{sec:2}
\subsection{Audits of AI systems and auditability}
Technical, legal, and ethical reasons mandate audits of AI systems to ensure accountability for accurate, compliant, and fair systems \cite{Raji.2020, Falco.2021, Ayling.2022}. For this purpose, auditing techniques, for example, from finance, are adapted for AI systems \cite{Mokander.2022}. However, there are no standards for audit quality \cite{Alagic.2021}, and residual risks usually remain uncertain \cite{Knechel.2013}.

An audit analyses if an AI system complies with legal regulations, organizational standards, or ethical values. It compares claims made by stakeholders, like developers or deployers of AI-based LA tools, to the system's actual behavior. Claims concern an AI-based LA system's validity, utility, and ethics: i.e., its models, components, data sets, or scopes accurately and reliably measure what they intend to measure, and are suitable for the intended purpose (validity); how useful and effective the system is in real-world fulfilling its application (utility); ethical, moral, or legal standards are considered and satisfied (ethics) \cite{Minkkinen.2024, Ayling.2022}. AI's functionality, application field, and associated risks require interdisciplinary competencies and skills in mandating and conducting audits \cite{Landers.2023}. Auditors recover “auditable artifacts” \cite{Ayling.2022} to validate whether an AI system is implemented and operating as claimed. AI systems' design and implementation principles affect audits and processes of assessing claims, actual behavior, and auditable evidence \cite{Li.2024, Ayling.2022, Falco.2021}.

Auditability is given \textit{when a system is reviewable independently} \cite{Williams.2022, Wolnizer.2006}. \cite{Weigand.2013} conceptualize auditability as a) the system provides information on how relevant values should be used or produced (\textit{claims}), b) the system generates information on how relevant values are used or produced (\textit{evidence}), and c) stakeholders can \textit{validate} these claims based on the provided evidence. The complexity of AI systems creates special requirements for audits and, thus, for a system's auditability \cite{Minkkinen.2022}. \cite{Li.2024} propose a framework for auditability that focuses on the entire life cycle of AI systems, including training data, models, and organizational governance processes. \cite{Berghoff.2022} approach auditability of AI systems from a cyber security perspective where increased system complexity impedes system auditability. \cite{Raji.2020} propose a joint internal audit process of auditors and auditees, who provide claims and artifacts as evidence, and create a remediation plan to mitigate risks \cite[]{Raji.2020}. 

Claims and suitable evidence to validate claims are cornerstones of auditable systems. In practice, these usually are not readily available. Various challenges and limitations exist, decreasing the auditability of AI-integrating LA systems at all stages \cite{Simbeck.2023, Fernsel.2024}. 

\paragraph{Formulating claims}
\textit{Claims} are normative statements on a system’s functionality, scope, and purpose. System providers and system deployers define claims in system standards, targeted fields of application, scope and use cases, or as part of documentation \cite{Stoel.2012}. Other claims stem from ethical or moral standpoints, laws, regulations, and standards that guide software implementation and use \cite{Brundage.2020}.

The first challenge for audits is defining verifiable claims about a system's validity, utility, and ethics. Ideally, these are quantifiable and measurable \cite{Felderer.2021}, and provided by the audited system provider \cite{Brundage.2020, Raji.2020}. For example, an adequate utility claim for an AI-based LA system could be that it reduces the drop-out risk for students of a specific course by $x\,\%$. Or that a specific ML-model based on student behavior data accurately predicts students at risk to fail. More difficult would be the claim that a system is robust to biased data from underrepresented groups. Quantifiable performance indicators are usually easier to assess than ethical or moral values. In practice, especially ethical claims remain vaguely defined by vendors or deploying organizations, making auditors responsible for their operationalization \cite{Kochling.2020}. Although several auditing frameworks have been suggested (e.g., \cite{Raji.2020}), initiatives sought to standardize AI system audits \cite{Soler.2023, AlaPietila.2020, SAI.2023}, auditors still are missing harmonized standards for AI audits \cite{EC.2022, Mokander.2021}. Without guidelines, auditors must define and decide how to measure ethical values individually and subjectively \cite{Rzepka.2022, Slade.2019, Landers.2023}. Auditors are at risk of introducing their own biases into audits. For example, the definition of fairness may vary depending on context (i.e., equality or equity of the AI-based LA interventions). Fairness definitions also may be mutually exclusive \cite{Jacobs.2021}. Further, intersectional dimensions of inequality could neglected subsets of groups and introduce additional bias, e.g., when only gender and ethnicity are considered, but the social background is overlooked. For instance, auditors may assign test cases—individuals—to the wrong group or miss underrepresented groups \cite{Baker.2021, Suresh.2021}.

\paragraph{Collecting evidence}
For AI-based LA, \textit{evidence} is “relevant information about its execution” \cite{Alhajaili.2019} that allows us to analyze and trace errors. Auditees should enable evidence collection by organizational structures and processes that document a system’s operation, while developers should provide accessible AI systems \cite{Awwad.2020, Stoel.2012}.

Evidence proves or rejects derived claims. However, sometimes, neither the system nor its raw sources (program code, model weights, data used for training and testing) are accessible to auditors, e.g., for proprietary HR software \cite{Raghavan.2020} or predictive policing tools \cite{Alikhademi.2022}. In such cases, auditors must simulate algorithms and models to conduct data-based audit methods that assess the fair treatment of inputted data \cite{Alikhademi.2022}. This limits the audit’s effectiveness and is additionally aggravated by lacking balanced test data that represents marginalized groups \cite{Fernsel.2024}. Independent code-based audit methods are not feasible for proprietary systems, either. Lastly, the documentation of AI-integrating systems—proprietary or open—is often incomplete \cite{Alikhademi.2022, Tagharobi.2022, Raghavan.2020}.

\paragraph{Verifying claims}
Auditors with system access and evidence can verify whether an AI-integrating system meets the derived claims. However, AI-based LA systems present challenges when designing test cases and selecting test data. AI can handle a broad range of input data and assume more possible states than common software \cite{Berghoff.2022, Minkkinen.2022}. Therefore, a diverse and substantial amount of test data is required \cite{Tao.2019, Fernsel.2024}. Further, pre-deployment audits might not cover every possible use case and differ from de facto operational use \cite{Tao.2019}. After deployment, models can be updated by learning from new training data or through feedback loops, and introduce bias in the process \cite{Berghoff.2022, Felderer.2021, Awwad.2020}. Therefore, tests of an AI are not necessarily realistic and make continuous auditing necessary \cite{EitelPorter.2021, Awwad.2020, Mokander.2021, Minkkinen.2022}. 

When verifying claims, auditors interpret the obtained output of a running AI-based LA system, which is challenging because AI-integrating systems have more complex and interrelated components than regular software \cite{Minkkinen.2022}. Complex AI algorithms are often considered “black boxes” that require explanations of the model output to users or auditors \cite{Brundage.2020, Guidotti.2018}, which must be actively enabled as well.

\subsection{Enabling auditability of AI-based LA systems}
Because of the limited auditability of AI-integrating systems, some audits rely on self-audits and require auditees to answer questions about the design principles and guarantees to functionality and compliance \cite{Raji.2020}. While this is a valid approach, we argue that AI-based LA systems must be designed with auditability in mind to enable independent audits. Even though such systems are complex, system providers and deploying institutions can take steps to enable independent auditability. Auditable AI-based LA systems require planning, documentation, the implementation of specific functionalities, such as logs, APIs, monitoring tools or explanations. Further, access to the system sources, such as program code, model configuration, or data facilitates realistic auditing under field conditions.

The following suggestions concern system development and responsible institutions deploying AI-based LA. On the one hand, LA's auditibility requires systematically incorporated auditability principles early on in the development process; on the other hand, auditability requires institutional settings and information systems with well-documented organizational and technical processes, a clear scope of system utility, and openness to third-party auditors.

\paragraph{Planning for auditability}
Sufficient auditability will only be reached if it is planned for during the system design process. To help the completeness of evidence, “accountability plans” outline what and how information should be captured \cite{Naja.2022}. Workflows to increase auditability include logging model training and validation results, storing model metadata, and continuous monitoring \cite{Kreuzberger.2022, Minkkinen.2022}. Plans should also determine applicable definitions of ethical standards and data for their evaluation to account for vague operationalizable claims \cite{GaldonClavell.2020, Slade.2019, kitto2019practical}. Based on the accountability plan, institutions can adjust organizational processes for auditability (project and risk management, design and development processes).

\paragraph{Documentation}
Auditability is further influenced by the completeness of documentation. The AI Act requires documentation for high-risk AI systems on the system in general, the models, and the relevant data \cite{EuropeanUnion.2024}.

System-related documentation includes functionality and limitations of the system \cite{EuropeanUnion.2024}. Auditability can be increased by documenting design and implementation choices, policies, external requirements, and organizational processes like project and risk management \cite{Raji.2020, Jentzsch.2019, SAI.2023}. Examples of the documentation of a system design process can be found in \cite{ahn2021co, veljanova2023operationalising}. In addition, the scientific basis behind learning analytics systems should be clearly stated to justify their utility and validity.

Model-related documentation includes information on algorithms for training, testing, and validation, and model parameters. It should elaborate on the model performance (\cite{SAI.2023, Mitchell.2019}; \cite{EuropeanUnion.2024}), which includes complex evaluation metrics like the ROC curve or a model-specific “measure of confidence” for each output \cite{Ashmore.2022}. Further, model cards can provide helpful information to assess a model's capabilities, limitations, and ethical considerations \cite{Mitchell.2019}.

Data-related documentation contains the data structure \cite{SAI.2023, EHLEG.2019} and information on data provenance, including the data acquisition method, data transformations, and data processing \cite{SAI.2023, Gebru.2021}. Documenting provenance contributes to reproducibility and helps to discover where biases originate and which data operations (e.g., data processing steps) influence them \cite{Toreini.2022}. Aspects of data quality \cite{EuropeanUnion.2024} such as the balance of classes in the training, validation, and verification data sets \cite{SAI.2023}, and data completeness \cite{Ashmore.2022} should be documented as well. For data sets, documentation standards are Data Sheets \cite{Gebru.2021} and Data set Nutrition Labels \cite{Holland.2018}. \cite{jones2019reconsidering} emphasized the importance of integrating data documentation into LA research, specifically data on students is operationalized and interpreted. During the documentation phase, major ethical concerns due to poor communication of institutional principles or policies arise, which are intransparent about benefits to the institution, and which offer no option for students to opt out \cite{Alfredo.2024}.

\paragraph{Providing sources}
AI-based LA results are often hard to reproduce \cite{haim2023open}. Auditability can be increased by providing raw sources of the system for evaluation purposes, including its source code, models, model weights and the training and test data \cite{Tagharobi.2022, SAI.2023}. Privacy issues may prohibit access to raw data. For such cases, auditors could collect or create (synthesized) test data  \cite{ElEmam.2020}. Additionally, synthetic data can be helpful when data for underrepresented minorities is scarce \cite{ElEmam.2020}.

\paragraph{Implementing auditability}
Enabling auditability of AI-based LA requires specific system functionalities for making system information accessible for auditors, such as secure system access  \cite{Awwad.2020}, logging \cite{EitelPorter.2021, AlaPietila.2020, Bose.2019},  monitoring tools \cite{EitelPorter.2021, Alhajaili.2019}, and explanations for model behavior \cite{Shneiderman.2020, Guidotti.2018}.  

 Secure system access for external auditors, e.g., via APIs, is a prerequisite for an audit \cite{Awwad.2020, WH.2022}. \cite{Springer.2019} show that APIs allow systematic tests of scenarios based on the system’s claims. APIs can also enable secure third-party access to logs \cite{Alla.2021}. Auditors can use logs to understand data flows \cite{Falco.2021}, or the recorded production process of predictions, data sets, and results \cite{Kale.2022}. Therefore, logs enhance the auditability of AI systems \cite{Brundage.2020, Shneiderman.2020}. Monitoring tools help to analyze performance, detect model behavior changes, and recognize violations of (ethical) constraints \cite{EitelPorter.2021}. They track various aspects of a system, such as model input, the environment of use, internal model properties, and model output \cite{Ashmore.2022}. Post-market monitoring after deployment is a constant part of the AI life cycle \cite{Alla.2021, Minkkinen.2022} and required under the AI Act \cite{EuropeanUnion.2024}. Explainable AI assists in making sense of models and their output. Feature importance or counterfactual explanations are part of the user interface and can help users or auditors to understand and judge, for example, what features contribute to classifications of students as \enquote{at risk to fail} \cite{Shneiderman.2020}.


\section{A framework for assessing auditability of AI systems}
\label{sec:3}

\begin{figure}[bt]
  \centering
  \includegraphics[width=0.4\linewidth]{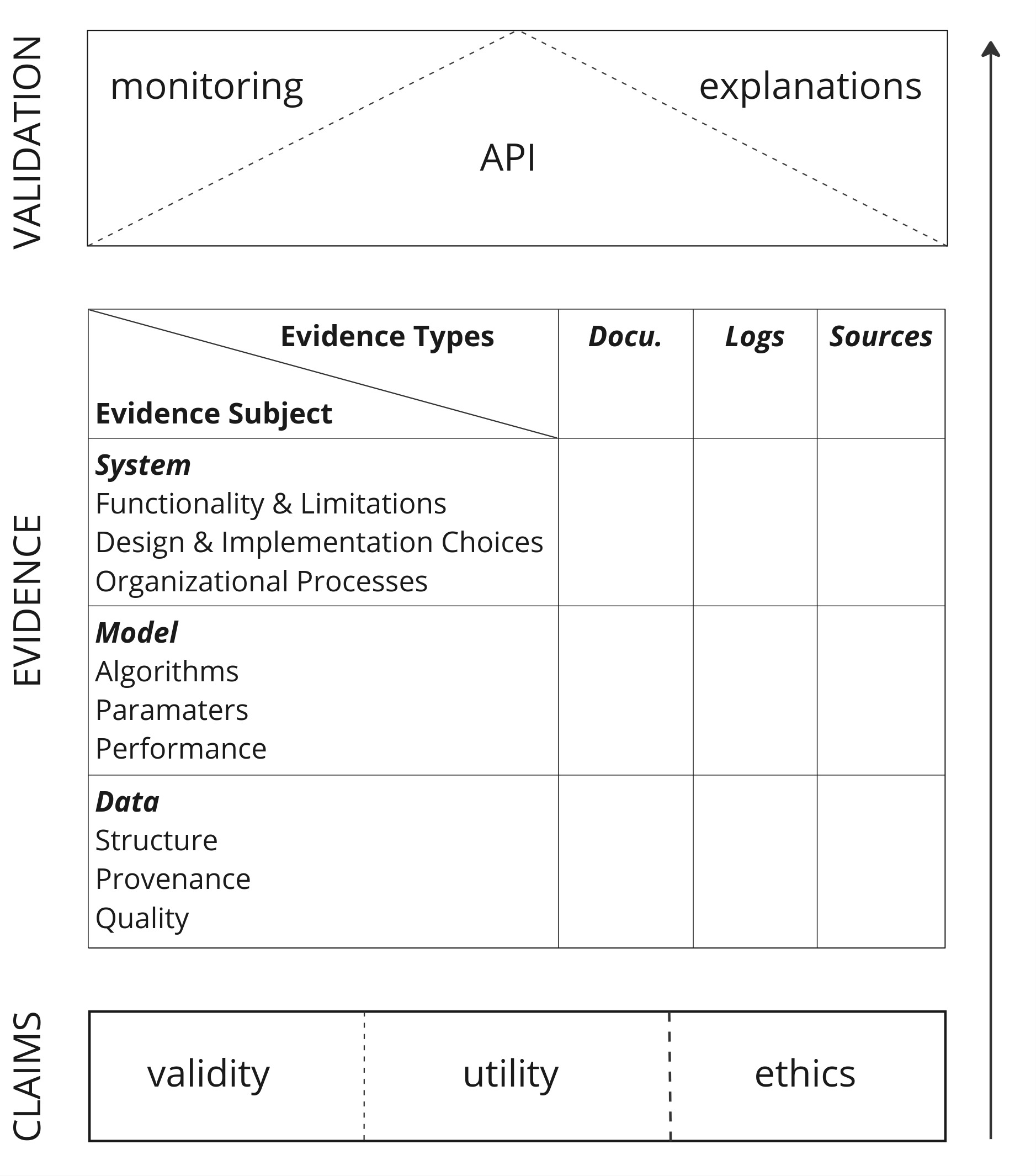}
  \caption{A framework for assessing auditability of AI systems}
  \label{fig:framework}
\end{figure}

Based on the discussion of audits and auditability of AI systems and methods to enhance the auditability of AI-integrating systems, we conceptualize a framework to assess auditability of AI-based LA systems and identify opportunities for their improvement. Figure~\ref{fig:framework} visualizes the framework. Any audit process has three steps displayed from the bottom to the top: first, auditors derive verifiable claims about the system, then identify, generate, and collect suitable evidence, and finally validate the claims based on the evidence.

\paragraph{Verifyable claims}
Developers or deploying organizations ensure the properties of the AI-based LA system and the processes in which it is applied. Auditors can derive verifiable claims from such assurance statements, which form the benchmark for the actual functioning of AI-based LA \cite{Brundage.2020}. Claims concern \textit{validity}: are methods correctly applied in the system, and is the system output correct? \textit{Utility}: is the system’s functionality helpful in its use case? Finally, \textit{adherence to underlying ethical principles}: does a system comply with applicable law (GDPR) or latent social, organizational, or societal norms, e.g., corporate culture, accessibility, or diversity, equity, and inclusion (DEI) goals? \cite{Landers.2023}

\paragraph{Evidence}
Once auditors define claims, they identify, create, and collect evidence \cite{Raji.2020}. Evidence comprises different subjects: system, model and data; and can take various forms as documentations, raw sources like source code, model weights and raw data, or logs \cite{Raji.2020, Brundage.2020, Tagharobi.2022, SAI.2023, EuropeanUnion.2024}.

Different evidence subjects assess specific aspects of AI-based LA systems. For the \textit{system}, evidence should prove the system’s functionality and its limitations; evidence must legitimize the underlying design and implementation choices and organizational processes \cite{Jentzsch.2019, SAI.2023, Raji.2020, EuropeanUnion.2024}. Evidence for the \textit{implemented models} concerns algorithms in use, model parameters, and model performance indicators \cite{SAI.2023, Mitchell.2019, EuropeanUnion.2024}. Evidence on \textit{data} informs about structure, provenance, and quality of test, training, or production data \cite{SAI.2023, Ashmore.2022, Gebru.2021, Toreini.2022, EuropeanUnion.2024}.

\paragraph{Means of validation}
In the validation step, auditors access and assess evidence to validate the claims about the AI-based LA system's validity, utility, and ethics. Means to validate can be integrated into the system---either as an interface to access raw data via APIs for further testing, in the form of monitoring tools to observe system output and parameters and deliver readily interpretable (real-time) results or as explainable AI principles on user interfaces \cite{Fernsel.2024, WH.2022, EitelPorter.2021, Bharadhwaj.2021, Minkkinen.2022}. 

\paragraph{Utilizing the Framework}
The framework aims to assess the auditability of an AI-based LA system and facilitate the design of auditable systems.  The identified claims and their specificity dictate what evidence subjects (system, model, data) and evidence types (sources, documentation, logs) are necessary to validate them. The evidence types, in turn, specify the technical means of validation. Auditors judge whether the evidence is sufficiently available and accessible. 

Gathering claims requires a heuristic search, document analysis, and Q\&A-interviews with responsible positions to determine the claims' relevance and hierarchy. Collecting evidence depends on subject and type, and is closely related to technical validation means. The most important type of evidence is documentation since it is the most accessible type to obtain and understand \cite{SAI.2023}. Arguably, the most challenging evidence could be source codes or logs of proprietary or security-sensitive systems \cite{Alikhademi.2022, Raghavan.2020}. However, these forms of evidence may be necessary to complete information from the documentation or establish credibility \cite{SAI.2023}.

In practice, the framework can be used to define ex-ante responsibilities in the audited organization for providing claims and evidence. It can be operationalized as a checklist to control the auditability of a system as an initial audit step or in the development cycle of a system. Since system development is an ongoing process, the framework assists in assessing the auditability on the developers' side and offers guardrails for quality assurance measures.

In the following two sections, we demonstrate our framework's practical application and utility to assess the auditability of AI-based LA systems. We assess the auditability of Moodle, a well-renowned tool with an integrated student dropout prediction system, and a research prototype tool that predicts student drop-out. The tools are AI-based LA that are potentially “high-risk” under the AI Act. Dropout prediction models have repeatedly been proven to work better for majority groups better represented in training data \cite{Gardner.2019, Rzepka.2022}. Therefore, there is a risk that some groups of students benefit less from the AI-based LA module than others. 

\section{Case study I: Auditability of Moodle’s student dropout prediction system}
\label{sec:4}

We chose Moodle because it is a commonly used open-source learning management system. Moodle’s dropout prediction system aims to prevent students from dropping out of a course \cite{Monllao.2018}. The software ships with an un-trained machine learning model (a model configuration) that, once trained on a particular Moodle platform, predicts whether a student is likely to drop out of a course \cite{Moodle.Documentation, Monllao.2018}. A model configuration can be tested in an “evaluation mode” before going live \cite{Moodle.Documentation}. Table~\ref{table1} lists the derived claims, and table~\ref{table2} summarizes the results of necessary and available claims.

\begin{table}[hbt]
 \caption{Claims made for Moodle dropout prediction system by type. \textbf{\textcolor{blue}{Bold blue}}: sufficient evidence available across types and subjects to validate claim; else: insufficient evidence to verify claim.}
 \centering
 \small
    \begin{tabular}{p{0.1\textwidth}p{0.8\textwidth}}
    \toprule
    \textbf{Type}       & \textbf{Claim}                                                    \\
    \toprule
    \textbf{Validity}   &   (v1) sufficiently good predictions                                              \\ 
                        &   (v2) cognitive depth and social breadth are valid indicators                                          \\
                        \cmidrule(lr){1-2}
    \textbf{Utility}    &   (u1) reduced dropout in online courses                              \\
    \cmidrule(lr){1-2}
    \textbf{Ethics}     &   \textbf{\textcolor{blue}{(e1) AI-created predictions are marked as such}}                                             \\
                        &   \textbf{\textcolor{blue}{(e2) stakeholders can decide whether to use the system}}                                  \\
                        &   \textbf{\textcolor{blue}{(e3) GDPR conformity}}                                                                   \\
                        &   (e4) equal efficiency for learners of different locations and financial backgrounds \\
    \bottomrule
    \end{tabular}
 
    \label{table1}
\end{table}

\begin{table}[hbt]
 \caption{Overview of desirable and available evidence for each claim by evidence type and subject. \checkmark: evidence of type and subject is fully available. $\square$: incomplete evidence of type and subject. \textbf{\textcolor{blue}{Bold blue}}: sufficient evidence available across types and subjects to validate claim; else: insufficient evidence to verify claim.}
 \centering
 \small
    \begin{tabular}{p{0.2\textwidth}p{0.2\textwidth}p{0.2\textwidth}p{0.2\textwidth}}
    \toprule
    \textbf{Evidence Type}  & \multicolumn{3}{c}{\textbf{Evidence Subject}} \\
    \toprule
    
                  &  \textbf{Documentation}  & \textbf{Logs}    & \textbf{Sources}  \\
                     \cmidrule(lr){2-4}

    \textit{System}                             & \cellcolor{lightgray}                         & \cellcolor{lightgray}                 &   \cellcolor{lightgray} \\
    Functionality \& Limitations                &   \textbf{\textcolor{blue}{e1\checkmark}}, 
                                                    \textbf{\textcolor{blue}{e2\checkmark}}, 
                                                    \textbf{\textcolor{blue}{e3$\square$}}, 
                                                    e4$\square$                                 &                                      &    \textbf{\textcolor{blue}{e3\checkmark}}, 
                                                                                                                                            e4\checkmark  \\
    Design \& Implementation choices            &   v1$\square$, 
                                                    v2$\square$, 
                                                    u1$\square$, 
                                                    \textbf{\textcolor{blue}{e3$\square$}}, 
                                                    e4$\square$                                 &                                       & \cellcolor{lightgray} \\
    Organizational processes                    &   v1\checkmark, 
                                                    u1$\square$, 
                                                    \textbf{\textcolor{blue}{e3$\square$}}, 
                                                    e4$\square$                                 &                                       & \cellcolor{lightgray} \\
    \cmidrule(lr){1-4}
    \textit{Model}                              & \cellcolor{lightgray}                         &  \cellcolor{lightgray}                &  \\
    Algorithms                                  &   v1\checkmark, 
                                                    \textbf{\textcolor{blue}{e3$\square$}}, 
                                                    e4$\square$                                 &                                       & \textbf{\textcolor{blue}{e3\checkmark}}, e4\checkmark \\
    Parameters                                  & v1$\square$, e4$\square$                      &                                       & v1\checkmark, e4\checkmark \\
    Performance                                 & e4$\square$                                   & v1$\square$, u1$\square$, e4$\square$ & \cellcolor{lightgray} \\
    \cmidrule(lr){1-4}
    \textit{Data}                               & \cellcolor{lightgray}                         &  \cellcolor{lightgray}                & v1$\square$, v2$\square$, u1$\square$, e4$\square$ \\
    Structure                                   & v1\checkmark                                  &                                       &  \cellcolor{lightgray} \\
    Provenance                                  & v1\checkmark, v2$\square$, e4$\square$        & v2$\square$, e4$\square$              &  \cellcolor{lightgray} \\
    Quality                                     & v1$\square$, e4$\square$                      &                                       &  \cellcolor{lightgray} \\
    \bottomrule
    \end{tabular}
   
    \label{table2}
\end{table}

\subsection{Claims}
The first step when verifying auditability is determining which claims concerning validity, utility, and compliance to ethical norms are made \cite{Landers.2023}. Therefore, we consult the documentation of Moodle’s student dropout prediction system and additional literature. 

These sources reference two validity-related claims (e.g., are methods correctly applied in the system, and is the system output correct?). The first validity-related claim \textit{v1} is that “[t]he accuracy and recall of the presented prediction model for predicting at-risk students are good for a production system” \cite{Monllao.2018}. Additionally, since the dropout prediction model design is based on \cite{Garrison.1999}'s “Community of Inquiry” framework, a second claim \textit{v2} is that cognitive depth (metric applied in Moodle for “cognitive presence” of a student) and social breadth (metric applied in Moodle for “social presence” of a student) are valid indicators for dropout prediction \cite{Moodle.Documentation}.

Furthermore, broad references can be found to the utility (is the system functionality useful in its specific context?) of the dropout prediction system. Moodle’s LA system should “not only predict events, but change them to be more positive” \cite{Moodle.Documentation}. The Moodle documentation asserts that the dropout prediction system is most useful for courses that run entirely online due to features that rely on Moodle activities \cite{Moodle.Documentation}. As utility claim \textit{u1}, we can formulate that the dropout prediction system reduces dropout rates in online courses.

MoodleHQ, the organization behind Moodle, explains which ethical principles drive the implementation and use of AI in Moodle: Users should always know when AI is used, stakeholders should be able to decide which AI components to use, AI components should preserve users’ data privacy and security, and AI components should be efficient for all learners, “regardless of their location or financial situation” \cite{Moodle.AIprinciples}. Four ethics-related claims can be derived from these principles. Firstly, dropout predictions are marked as results of an AI (\textit{e1}). Secondly, stakeholders can decide whether to use the dropout prediction system (\textit{e2}). This implies that institutions can activate the feature, and learners can opt in or out. Third, student data collection, processing, and storage by the dropout prediction system follows the EU General Data Protection Regulation GDPR (\textit{e3}). And lastly, the dropout prediction model is equally efficient for all learners, regardless of their geographical location and financial background (\textit{e4}).

\subsection{Required evidence}
We established that evidence in the form of documentation, raw sources and logs is suitable to verify claims. Evidence can concern aspects of Moodle’s dropout prediction system (the system), the dropout prediction model, and the underlying data. This subsection examines which evidence is required to validate which claim. The claims and evidence subjects are indicated in cursive. 

\paragraph{Validity}
To prove \textit{v1} (sufficiently good predictions), the most reliable way would be to reproduce the quality assessment conducted by \cite{Monllao.2018}. To evaluate the dropout prediction configuration, an auditor requires access to a Moodle system with test data \textit{(data sources)} to calculate the \textit{model performance}. We call this type of system a “test system”. In the absence of openly available test data, additional documentation on \textit{data quality} requirements is helpful to acquire suitable test data. In this use case, data can only be acquired by exporting data from a Moodle platform, not through synthesis. This is because of the lack of seed data. Even if sufficient information on data properties was available, auditors cannot import data for model input, because the model requires meaningful related data, which cannot be synthesized \cite{Fernsel.2024}. 

If suitable test data cannot be obtained to reproduce MoodleHQ’s quality assessment, the reliability of the evaluation conducted by \cite{Monllao.2018} can at least be judged. The auditor needs to know the details of the quality evaluation \textit{(organizational processes)} and the properties of the used training and test data \textit{(data structure, provenance, and quality)}. Additionally, information on \textit{model algorithms} (training and testing, including feedback loops) and which \textit{model parameters} were chosen and why \textit{(system design and implementation choices)} could help to identify erroneous implementations that lead to biased results.

To prove \textit{v2} (cognitive depth and social breadth are valid indicators), auditors need to verify whether this claim is scientifically supported by studies. This information can be expected in the documentation on the application of the “Community of Inquiry” framework into cognitive depth and social breadth indicators \textit{(system design and implementation choices)}. Trust can be increased by examining the importance of each indicator on the predictions made on a “production systems”---a Moodle instance that is using the dropout prediction model \textit{(data provenance)}.

\paragraph{Utility}
The utility-related claim (reduced dropout in online courses) requires evidence that indicates the system’s impact on student behavior in online courses. This information could be found in the documentation on the scientific foundation of the dropout prediction system and in conducted studies \textit{(design and implementation choices)}, as well as in applied evidence-based design methods \textit{(organizational processes)}. If such evidence is unavailable, auditors may verify the system’s utility by analyzing the feedback given by humans for predictions; i.e., did a student drop out and did Moodle predict this correctly \textit{(model performance)}—provided that they have access to a production system. 

\paragraph{Ethics}
To validate \textit{e1} (AI-created predictions are marked as such) and \textit{e2} (stakeholders can decide whether to use the system), documentation on \textit{system functionality and limitations} can be helpful. To assess \textit{e3} (GDPR compliance) documentation on \textit{system functionality and limitations}, \textit{design and implementation choices}, \textit{algorithms} and \textit{organizational processes} could show whether data privacy and security mechanisms have been included. Analyzing the source code could provide detailed information about the system's actual behavior compared to potentially incomplete documentation.

Several pieces of evidence can help verify \textit{e4} (equal model performance across groups). The documentation on \textit{system functionality and limitations}, \textit{design and implementation choices}, and \textit{organizational processes} could reveal structural issues that might lead to a biased system. It could also contain information on how risks are handled. Evidence on the \textit{model algorithms} and \textit{parameters} can uncover further ethical issues \cite{Tagharobi.2022}. Documentation of the \textit{model performance} by risk group could indicate the equality of prediction for different groups. Properties of the training and test data \textit{(data provenance, data quality)} must be known to ensure the validity of the performance evaluation. Trust can be increased if \textit{data} is available for reproducing or extending quality measurements.

Auditors can examine a production system's user interface and functionalities if the documentation for any ethics-related claims is incomplete. Specifically, a quality assessment of the dropout prediction model per group could provide reliable evidence on the fairness of the dropout prediction model (\textit{e4}). However, this requires access to data on the relevant user properties, which is not given (\textit{their geographical location and financial status}).

\subsection{Means of validation}
In this subsection, we assess to what extent Moodle’s dropout prediction system implements interfaces to access and collect evidence for validating claims.  

\paragraph{API}
Moodle does not provide an API for secure third-party access to the dropout prediction system. However, the internal “Analytics API” may be used to access and extend the machine learning capabilities of Moodle with a plugin \cite{Monllao.2018}. In a different publication, we successfully used this approach to increase the auditability of Moodle \cite{Fernsel.2024}.

\paragraph{Monitoring}
The primary monitoring capability is the “evaluation mode” for evaluating model configurations or models trained on other Moodle instances \cite{Moodle.Documentation}. An auditor needs access to a test system to use this monitoring capability. The “evaluation mode” trains a new model on a data split from finished courses and tests it against the remaining data \cite{Moodle.Documentation}. A model trained on a different platform is evaluated by testing it against the data on the new platform. When doing so, the model trained for evaluation is not retained. The “evaluation mode” returns two values per selected analysis interval: the weighted F1-score and the standard deviation \cite{Moodle.Documentation}. Depending on the chosen machine learning module, more values like the Matthews’ correlation coefficient may be returned \cite{Monllao.2018}. 

Smaller monitoring capabilities for production systems are also available. Auditors can monitor which courses cannot be used by the model, which students have been classified as at risk of dropping out, which indicators have been calculated for which student, and which human feedback has been given for the dropout prediction model: correct, “not applicable” or “incorrectly flagged” \cite{Moodle.Documentation}. 

\paragraph{Explanations}
Moodle integrates explanations for AI outputs in production systems. As mentioned above, Moodle monitors the results of the dropout prediction model together with the calculated indicators per student. Moodle highlights influential indicators to explain the model result \cite{Moodle.Documentation}.

\subsection{Evidence accessibility}
Finally, we analyze whether the required evidence is sufficiently available and accessible for validating the claims. The claim and evidence subject are indicated in cursive.

\paragraph{Validity}
To prove \textit{v1} auditors should reproduce the dropout prediction model performance assessments from \cite{Monllao.2018}. The performance assessment requires a test system to obtain logs of the \textit{model performance}. However, as mentioned above, \textit{data sources} for the test system are not publicly available. The auditor must acquire suitable test data to set up a test system. Quality requirements for the acquired data are available \cite{Moodle.Documentation, Monllao.2018} (\textit{data quality}).

If no test system is available, the validity of the quality evaluation may be estimated by reviewing information on the properties and production of the system: \textit{organizational processes, data structure, data provenance, data quality, model algorithms, model parameters} and \textit{system design and implementation choices}. Basic information on the \textit{structure and provenance} of the data used for MoodleHQ’s quality assessment describe \cite{Monllao.2018}. They also elaborate on the data quantity. The model training algorithms—logistic regression and a feed-forward neural network—\textit{(model algorithm)} and the model evaluation methods are documented as well \textit{(organizational processes)}. Furthermore, a possible feedback loop exists when prediction feedback is included in the training data for future models \cite{Moodle.Documentation} \textit{(model algorithms)}. The configurable dropout prediction \textit{model parameters} (analysis interval and context) are also documented \cite{Moodle.Documentation, Monllao.2018}. 

However, auditors cannot rely on evidence from the documentation alone to validate MoodleHQ's quality evaluation. They will need to analyze relevant parts of the source code: the values for fixed \textit{model parameters} (like the number of training epochs, learning rate, or batch size) are neither documented nor logged  and can only be found in the source code. Also, a source code analysis by \cite{Tagharobi.2022} found an undocumented 500MB limit for training data. Even by complementing information from the documentation with a source code analysis, some evidence that would help assess validity-related claims is missing. This concerns firstly, information on additional and more detailed \textit{data quality} aspects, such as data completeness or data quantity per student or course; and secondly, information on whether model parameters, such as learning rate and batch size, were considered and if so why they were discarded \textit{(system design and implementation choices)}. To conclude,   available and accessible evidence is insufficient to fully validate claim \textit{v1} that the dropout prediction model correctly predicts dropout risks. 

The documentation on applying the “Community of Inquiry” framework into cognitive depth and social breadth indicators \textit{(system design and implementation choices)} offers a starting point for auditors to validate \textit{v2} \cite{Moodle.Documentation, Monllao.2018}. MoodleHQ does not provide studies that support their indicator definitions. If an auditor can access a production system, she could review the explanations logged for the model’s predictions \textit{(data provenance)} and evaluate the soundness of the chosen indicators. Since this type of access could be challenging, we deem the available evidence insufficient to effectively assess claim \textit{v2} that cognitive depth and social breadth are valid indicators.

\paragraph{Utility}
To assess the utility-related claim \textit{u1}, the first step is to review the available documentation.
The scientific theory behind the choice of model features is explained thoroughly \cite{Monllao.2018}, but no studies on the model’s impact are documented \textit{(design and implementation choices)}. Project management, design, and development processes are not documented either \textit{(organizational processes)}. Production system-specific utility may be analyzed by viewing aggregated information about the feedback given by humans for predictions \textit{(model performance)}. In summary, the available evidence does not validate the claim \textit{u1} that the dropout prediction system reduces dropout rates in online courses. 

\paragraph{Ethics}
To validate ethics-based claim \textit{e1}, documentation on \textit{system functionality and limitations} can be reviewed. The screenshots in the Moodle documentation show that users viewing dropout predictions are made aware of the uncertainty of predictions \cite{Moodle.Documentation}. However, users do not appear explicitly informed that the AI-based LA system calculates the predictions. The available evidence allows us to reject claim \textit{e1} that AI-created predictions are marked as such. 

Documentation on \textit{system functionality and limitations} could also help validate claim \textit{e2}. The documentation shows that teachers can decide how to use the predictions, and administrators can turn the dropout prediction system on or off \cite{Moodle.Documentation}. Students do not appear able to opt in or out of being classified by the dropout prediction model. In conclusion, the available evidence indicates that claim \textit{e2}—stakeholders can decide whether to use the dropout prediction system—can only partly be confirmed.

Concerning \textit{e3}, the documentation on \textit{system functionality and limitations}, \textit{design and implementation choices}, and  \textit{organizational processes} does not explicitly explain what has been done to comply with the GDPR, except that exportable data is anonymous and access to insights can be managed \cite{Moodle.Documentation}. We conclude that only a source code analysis \textit{(algorithms, system functionality, and limitations)} can assess the claim that the dropout prediction system is GDPR compliant.

To validate \textit{e4}, documentation on the choice of model features and their underlying principles \textit{(design and implementation choices)}, as well as limiting technical factors \textit{(system functionality and limitations)} hint at existing or absent bias in the dropout prediction system \cite{Monllao.2018, Moodle.AnalyticsAPI, Moodle.Documentation}. However, they do not mention any known issues. Source code analysis is required to complement the documentation. However, it cannot rule out any bias \cite{Tagharobi.2022}.
Evaluating the \textit{model performance} per group could provide additional evidence on model fairness. Such a quality assessment is not documented and thus needs to be conducted by the auditor. Access to a production system (including the database) and \textit{data sources}, including demographic data, is necessary. The “evaluation mode” cannot conduct such a performance evaluation because it does not return ethics-related metrics (e.g., group-based performance metrics). However, the auditor could calculate these metrics from predictions, truth values, and user data in a production system. This data can be collected from a log on what was predicted, when, and for which user \cite{Green.2023}. While users can disclose information on their geographical location, Moodle does not collect information on their financial status \cite{Green.2023}. The documentation on \textit{system functionality and limitations} and \textit{design and implementation choices} \cite{Monllao.2018, Moodle.AnalyticsAPI, Moodle.Documentation} does not mention model fairness. No evidence could be found that the risk of model bias was considered in the design and development (\textit{organizational processes}). No information on the \textit{data provenance} (e.g., information on data acquisition and pre-processing) or relevant \textit{data quality} (e.g., information on representativeness) for MoodleHQ’s quality assessment is available \cite{Monllao.2018}. We conclude that insufficient evidence is available and accessible for an efficient audit of claim \textit{e4} that model performance is equally high across groups. 

\section{ Case study II: Auditability of a research dropout-prediction system }
\label{sec:5}

In this second, shorter case study, we apply the auditability assessment framework to the extended approach published in \cite{zhidkikh2024reproducing} based on \cite{van2023pass}. This research prototype represents an early stage of an AI-based LA system. The study implements a dropout prediction in an introductory computer science course using log data from the course and student's self-reported data. Table~\ref{table3} summarizes the established claims about the system’s validity, utility, and ethics. Table~\ref{table4} describes the available evidence and if it is sufficient to validate the claims.

\begin{table}[hbt]
 \caption{Claims made for Moodle dropout prediction system by type. \textbf{\textcolor{blue}{Bold blue}}: sufficient evidence available across types and subjects to validate claim; else: insufficient evidence to verify claim.}
 \centering
 \small
    \begin{tabular}{p{0.1\textwidth}p{0.8\textwidth}}
    \toprule
    \textbf{Type}       & \textbf{Claim}\\
    \toprule
    \textbf{Validity}   &   \textbf{\textcolor{blue}{(v1) strong dropout predictions}}\\ 
                        &   \textbf{\textcolor{blue}{(v2) student behavior and self-reported data enhance dropout prediction accuracy}}\\
                        \cmidrule(lr){1-2}
    \textbf{Utility}    &    \textbf{\textcolor{blue}{(u1) dropout prediction in CS education aids in detecting struggling students for early support}}\\
    \cmidrule(lr){1-2}
    \textbf{Ethics}     &    \textbf{\textcolor{blue}{(e1) framework and the models are transparent}}\\
                        &   (e2) privacy-first drop-out prediction \\
    \bottomrule
    \end{tabular}
    \label{table3}
\end{table}

\begin{table}[hbt]
 \caption{Overview of desirable and available evidence for each claim by evidence type and subject. \checkmark: evidence of type and subject is fully available. $\square$: incomplete evidence of type and subject. \textbf{\textcolor{blue}{Bold blue}}: sufficient evidence available across types and subjects to validate claim; else: insufficient evidence to verify claim.}
 \centering
 \small
    \begin{tabular}{p{0.2\textwidth}p{0.2\textwidth}p{0.2\textwidth}p{0.2\textwidth}}
    \toprule
    \textbf{Evidence Type}  & \multicolumn{3}{c}{\textbf{Evidence Subject}} \\
    \toprule
    
                  &  \textbf{Documentation}  & \textbf{Logs}    & \textbf{Sources}  \\
                     \cmidrule(lr){2-4}
    \textit{System}                             &                        &                &    \\
    Functionality \& Limitations                &   {v1\checkmark}, 
                                                   {v2\checkmark}, 
                                                    \textbf{\textcolor{blue}{u1\checkmark}}     &                                      &    {e1\checkmark} \\
    Design \& Implementation choices            &	u1$\square$, e2$\square$ &	&	\\
    Organizational processes                    &	u1$\square$, e2$\square$ &	&	\\                         
    \cmidrule(lr){1-4}
    \textit{Model}                              &  \textbf{\textcolor{blue}{v1\checkmark}}, 
                                                    \textbf{\textcolor{blue}{v2\checkmark}}, {e1\checkmark} &   & \textbf{\textcolor{blue}{e1\checkmark}}	\\
    Algorithms                                  & {e1\checkmark}   &	& \textbf{\textcolor{blue}{e1\checkmark}}	\\
    Parameters                                  &  \textbf{\textcolor{blue}{v2\checkmark}} &  & \\
    Performance                                 &  \textbf{\textcolor{blue}{v2\checkmark}} &   &	\\
    \cmidrule(lr){1-4}
    \textit{Data}                               &	&	&	\\
    Structure                                   &	u1\checkmark &	&  e2$\square$, \textbf{\textcolor{blue}{e1\checkmark}}	\\
    Provenance                                  &	v2\checkmark,  u1\checkmark, e2$\square$  &	&	\\
    Quality                                     &	v1$\square$, v2$\square$ &	&	\\
    \bottomrule
    \end{tabular}
   
    \label{table4}
\end{table}

\subsection{Claims}
With regards to the validity of the system, the authors of the system claim \textit{v1} that their models “provide strong dropout predictions” \cite{zhidkikh2024reproducing}. They also claim \textit{v2} that “combining student behaviour and self-reported data early in a course enhances dropout prediction accuracy”. Thus, they claim that self-reported student motivation and aptitude data are valid indicators. 

The authors claim the utility (\textit{u1}) of their approach is that “dropout prediction in computer science education aids in detecting struggling students for early support” \cite{zhidkikh2024reproducing}.

For ethics, they state the “framework and the models are transparent” (\textit{e1}); the system and its results are interpretable and accessible to human comprehension. Further, since \cite{van2023pass} is considered a privacy-first approach to student data in AI-based LA, their extended implementation shares the claim \textit{e2} of a privacy-friendly drop-out prediction system. 

In contrast to our Moodle case study, \cite{zhidkikh2024reproducing} make all claims about a specific course at a particular university in a specific period. While they are reproducing prior work by \cite{van2023pass} and are thus extending the approach’s generalizability, they do not claim that their model works in all courses at all points in time, as is the case with Moodle.

\subsection{Evidence}
Given the research nature of the project, the research article is an integral part of the system’s documentation and explains the \textit{system’s functionality and limitations}. Discussing the limitations of their approach, the authors argue that the results in the form of feature importance are not “explainable to educators” (\textit{e1}) and not yet comparable to a “simple-to-use dashboard” \cite {zhidkikh2024reproducing}.
 With regards to the system’s \textit{design and implementation choices} as well as the \textit{organizational processes} of system creation, the article states that the work aimed at reproducing and extending a prior system by \cite{van2023pass}.
\cite{zhidkikh2024reproducing} provide models and sources for the analysis in a Github-Repository. The code is documented in a \texttt{readme} file, which includes a section on setup and basic usage. 
They describe the data collection and pre-processing (\textit{data provenance}) as well as the model building in section~3 of their paper. No system logs are provided. The information, code, and data provided are sufficient to run the code. The \textit{Model performance (v1)} is discussed in detail in section~4 of the paper with a detailed analysis of the F1 score and balanced accuracy. To prioritize explainability (\textit{e1}), the logistic regression classifier is favoured over the random forest model as it provides feature weights. The authors also discuss the effect of model choice on accuracy and model drift effect if older data is used \cite{zhidkikh2024reproducing}. The validity of self-reported data (\textit{v2}) is underpinned by a comparison of model accuracy with and without self-reported data, leading to the conclusion that self-reported data is improving accuracy, especially in the first weeks of the course \cite{zhidkikh2024reproducing}.
The utility of early detection for supporting struggling students \textit{u1} is justified with scientific research in the paper's theoretical and concluding sections. The privacy-friendly implementation (\textit{e2}) is not explained further: it would require further information on \textit{design and implementation choices}, and \textit{organizational processes}.

\subsection{Means of validation}
The system presented in \cite {zhidkikh2024reproducing} does not provide APIs or monitoring tools. Expert users can, however, consult feature weights of the logistic regression model, which is inherently explainable. The research prototype system uses the command line interface and Jupyter notebooks. Both user interfaces can be considered to address expert users and not educators in general. It is questionable if auditors are in the scope to assess such tools. However, given the prototype state, auditability can serve the developers to test and refine their tool to validate their claims about the system’s functionality.

\subsection{Auditability}
The system documentation, as provided in the research article, and the open-source code are sufficient to ensure the auditability of the system concerning the claims made by authors. This underlines the importance of making reproducible code and data accessible in scientific papers.
The documentation lacks a description of the data structure outputted by the learning environment. This limits the auditability of the claims to the exact scenario described in the paper. To ensure auditability for broader, more generalized claims, more detailed documentation of implementation and data would be necessary. 


\section{Discussion}
\label{sec:6}

We have demonstrated that our auditability assessment framework is helpful for AI-based LA systems by successfully applying it to Moodle’s dropout prediction feature and a prototype dropout prediction system. Through the structured approach of analysing claims, evidence, and means of validation, we support auditors to systematically approach complex AI-based systems (\textbf{RQ1}).

Although Moodle is open source, well-documented, and includes a comprehensive logging system with explanations, only three of seven identified claims are effectively auditable. The assessment of Moodle shows that future LA systems need to provide system access to third-party auditors, e.g., by creating “auditor” roles, recording data (anonymized training data, predictions), and enabling auditors to control evaluation parameters. More documentation is needed on system design and implementation choices to justify the validity and ethical design of the system. When documentation is incomplete or not trustworthy enough, additional evidence for an audit of the dropout prediction system must be collected from the system, e.g., by monitoring. Two significant challenges hinder this approach. First, Moodle’s model monitoring capabilities are inadequate. Predictions are not preserved when evaluating a model configuration and are inaccessible to the auditor. Thus, the auditor cannot verify the model’s performance and has to rely on minimal metrics returned by Moodle. Second, publicly available test data is lacking, which impedes data-based audits of Moodle’s dropout prediction model. However, complex systems like Moodle can benefit from the supporting feature of the framework for auditability and inform the development of suitable fixes and features that retrofit auditability \cite{Fernsel.2024}.

The assessment of the prototype LA shows that small-scale prototypes are less complex to audit, especially when their development and application are scientifically documented. This mainly boosts claims that are founded in scientific evidence. Potentially, their smaller scale makes claims and suitable evidence more manageable and may guide their development toward reflecting auditability early on (\textbf{RQ2}). Nonetheless, developing with auditability in mind increases complexity. A minimum set of claims could concern validity (accuracy, or performance) and ethical standards (privacy, GDPR-compliance, sensitivity for biased data) of an AI-based LA, while utility claims usually are tested in extensive piloting phases. 

However, limitations exist, some of which merit further research in the future to extend the auditability framework and add to its generalizability. The framework is only partly applicable to LA systems that don’t utilize AI. In our discussion of Moodle and a prototype AI-based LA, we could not account for the organizational setting, i.e., the institution where a running instance would be embedded. These institutions can provide additional claims, derived, for example, from organizational culture or corporate social responsibility statements. Further, organizations and institutions offer additional evidence to validate claims that may emerge in documented decision processes, legal statements, or technical scopes. These additional spheres must be included in any assessment of the auditability of AI-based LA systems and highlight that AI audits are complex and multifaceted practices that scrutinize technical and organizational aspects of AI implementation \cite{Landers.2023}. Further, types of AI-based LA are not distinguished: analytical, predictive, or predictive LA systems place different demands on their AI components. In particular, systems that are specialized in individual prediction or that recommend concrete action for individuals must be audited particularly rigorously. The more aggregated data is, the more lax audits---and therefore auditability---could be.

The auditability assessment framework requires time and effort---deriving claims and evidence, while technical means for validation might not be readily accessible. We are confident that development benefits from our auditability framework as will audits improve in quality. This justifies additional time and effort to procure more robust and ethically fair software for “high-risk” application fields.


\section{Conclusion}
\label{sec:7}

Despite the increasing demand for auditing AI systems, auditability is a neglected design requirement. This challenges independent audits with the lack of documentation \cite{Alikhademi.2022, Raghavan.2020, Tagharobi.2022}, restricted access to the system and its raw sources (code, model weights, or data) \cite{Raghavan.2020, Alikhademi.2022}, and incomprehensible system output \cite{Berghoff.2022, Felderer.2021, Mokander.2021}. Additionally, system-independent factors, such as heterogeneous ethical standards \cite{Mokander.2021} and difficulty in achieving test coverage for AI-integrating systems diminish auditability \cite{Berghoff.2022, Tao.2019}.

Following a review of auditability in general, AI audit challenges, and factors enabling AI auditability, we suggest a framework for a systematic approach to assess and ensure specific requirements for the auditability of AI-based LA. Our framework is based on three pillars: claims, evidence, and means of validation. To make AI systems auditable, system providers and deployers must provide certifiable claims about validity, utility, and ethics \cite{Landers.2023, Brundage.2020}. Depending on the claims and the audit procedure, substantial evidence must be made available to auditors: \textit{evidence types} include documentation \cite{Raji.2020, SAI.2023}, raw sources \cite{Tagharobi.2022, SAI.2023} and logs \cite{Brundage.2020}, \textit{evidence subjects} are the overall system, models and data \cite{EuropeanUnion.2024}. AI-integrating systems should provide APIs \cite{WH.2022, Springer.2019}, monitoring tools \cite{Ashmore.2022, Bharadhwaj.2021, EuropeanUnion.2024, EitelPorter.2021, Alhajaili.2019} and explanations \cite{Brundage.2020, Shneiderman.2020, Guidotti.2018} to enable the validation of evidence. Audit requirements and standards for AI audits are being developed. However, legislators and standardization bodies must consider auditability requirements as well. We see this as an important leverage point where our framework can be applied to derive process requirements for external audits, implement auditability by design in the QA of system development, and give stakeholders a way to insist on consistent audits. Finally, the framework supports developing and maintaining robust, trustworthy AI-based LA systems that foster acceptance among students and teaching professionals. 

We conclude that the proposed framework is useful for auditors and system providers to prepare for an audit and determine how much an AI-based LA system is auditable. Moreover, developers can benefit from the framework by identifying areas for improving the auditability of their products. We appeal to developers of AI-integrating systems to consider auditability right from the start when designing their systems to ensure trustworthy, ethical, and future-fit products that comply with current and upcoming legislation, such as the European AI Act. Considering that LA can potentially enhance learning outcomes \cite{Alfredo.2024}, increasing the auditability of LA systems ultimately leads to an improved learning experience for a broader audience.

\section*{Acknowledgments} 
The authors would like to thank several reviewers who improved the paper with their comments.

\section*{Declaration of Conflicting Interest} 
The author(s) declared no potential conflicts of interest with respect to the research, authorship, and/or publication of this article.

\section*{Funding} 
The underlying research was funded by the German Federal Ministry of Education and Research (BMBF) as part of the project “Investigating the fairness of learning analytics systems -- FairEnough; sub-project: Auditing of Learning Analytics Systems” (funding number: 16DHB4002).

\printbibliography

@article{Susnjak.2024,
 author = {Susnjak, Teo},
 year = {2024},
 title = {{Beyond Predictive Learning Analytics Modelling and onto Explainable Artificial Intelligence with Prescriptive Analytics and ChatGPT}},
 pages = {452--482},
 pagination = {page},
 volume = {34},
 journal = {{International Journal of Artificial Intelligence in Education}},
 %shorthand = {Beyond Predictive Learning Analytics Modelling and onto Explainable Artificial Intelligence with Prescriptive Analytics and ChatGPT},
 doi = {10.1007/s40593-023-00336-3},
 number = {2},
 abstract = {}
}

@misc{EuropeanUnion.2024,
 author = {{European Union}},
 year = {2024},
 title = {{Regulation (EU) 2024/1689 of the European Parliament and of the Council of 13 June 2024 laying down harmonised rules on artificial intelligence and amending Regulations (EC) No 300/2008, (EU) No 167/2013, (EU) No 168/2013, (EU) 2018/858, (EU) 2018/1139 and (EU) 2019/2144 and Directives 2014/90/EU, (EU) 2016/797 and (EU) 2020/1828 (Artificial Intelligence Act)}},
 url = {http://data.europa.eu/eli/reg/2024/1689/oj},
 number = {2024/1689},
 series = {{Regulation (EU)}},
 organization = {{Official Journal of the European Union}}
}

@misc{AlaPietila.2020,
 author = {Ala-Pietil{\"a}, Pekka and Bonnet, Yann and Bergmann, Urs and Bielikova, Maria and Bonefeld-Dahl, Cecilia and Bauer, Wilhelm and Bouarfa, Loubna and Chatila, Raja and Coeckelbergh, Mark and Dignum, Virginia and Gagn{\'e}, Jean-Francois and Goodey, Joanna and Haddadin, Sami and Hasselbalch, Gry and Heintz, Fredrik and Hidvegi, Fanny and H{\"o}ckner, Klaus and J{\'e}go-Laveissi{\`e}re, Mari-No{\"e}lle and K{\"a}rkk{\"a}inen, Leo and K{\"o}szegi, Sabine Theresia and Kroplewski, Robert and Martinkenaite, Ieva and Mallart, Raoul and Muller, Catelijne and Wendling, C{\'e}cile and O'Sullivan, Barry and Pachl, Ursula and Petit, Nicolas and Renda, Andrea and Rossi, Francesca and Yeung, Karen and Fogelman, Fran{\c{c}}oise Souli{\'e} and Tallinn, Jaan and Uszkoreit, Jakob and {van Wynsberghe}, Aimee},
 year = {2020},
 title = {{The Assessment List for Trustworthy Artificial Intelligence (ALTAI)}},
 publisher = {{European Commission}}
}

@inproceedings{Alhajaili.2019,
 author = {Alhajaili, Sara and Jhumka, Arshad},
 title = {{Auditability: An Approach to Ease Debugging of Reliable Distributed Systems}},
 pages = {227--2278},
 publisher = {IEEE},
 booktitle = {{2019 IEEE 24th Pacific Rim International Symposium on Dependable Computing (PRDC)}},
 year = {2019},
 address = {Kyoto, Japan},
 doi = {10.1109/PRDC47002.2019.00053}
}

@article{Alikhademi.2022,
 author = {Alikhademi, Kiana and Drobina, Emma and Prioleau, Diandra and Richardson, Brianna and Purves, Duncan and Gilbert, Juan E.},
 year = {2022},
 title = {{A review of predictive policing from the perspective of fairness}},
 pages = {1--17},
 volume = {30},
 number = {1},
 issn = {1572-8382},
 journal = {{Artificial Intelligence and Law}},
 doi = {10.1007/s10506-021-09286-4}
}

@incollection{Alla.2021,
 author = {Alla, Sridhar and Adari, Suman Kalyan},
 title = {{What Is MLOps?}},
 pages = {79--124},
 publisher = {Apress},
 isbn = {978-1-4842-6549-9},
 booktitle = {{Beginning MLOps with MLFlow: Deploy Models in AWS SageMaker, Google Cloud, and Microsoft Azure}},
 year = {2021},
 address = {Berkeley, CA},
 doi = {10.1007/978-1-4842-6549-9_3}
}

@article{Ashmore.2022,
 author = {Ashmore, Rob and Calinescu, Radu and Paterson, Colin},
 year = {2022},
 title = {{Assuring the Machine Learning Lifecycle: Desiderata, Methods, and Challenges}},
 pages = {1--39},
 volume = {54},
 number = {5},
 issn = {0360-0300},
 journal = {{ACM Computing Surveys}},
 doi = {10.1145/3453444}
}

@book{Awwad.2020,
 author = {Awwad, Yazeed and Fletcher, Richard and Frey, Daniel and Gandhi, Amit and Najafian, Maryam and Teodorescu, Mike},
 year = {2020},
 title = {{Exploring Fairness in Machine Learning for International Development}},
  address = {Cambridge},
 publisher = {{MIT D-Lab}}
}

@article{Baker.2021,
 author = {Baker, Ryan S. and Hawn, Aaron},
 year = {2022},
 title = {{Algorithmic Bias in Education}},
 pages = {1052--1092},
 volume = {32},
 number = {4},
 issn = {1560-4292},
 journal = {{International Journal of Artificial Intelligence in Education}},
 doi = {10.1007/s40593-021-00285-9}
}

@misc{Berghoff.2022,
 author = {Berghoff, Christian and B{\"o}ddinghaus, Jona and Danos, Vasilios and Davelaar, Gabrielle and Doms, Thomas and Ehrich, Heiko and Forrai, Alexandru and Grosu, Radu and Hamon, Ronan and Junklewitz, Henrik and Romanski, Simon and Samek, Wojciech and Schlesinger, Dirk and Stavesand, Jan-Eve and Steinbach, Sebastian and Wiegand, Thomas},
 year = {2022},
 title = {{Towards Auditable AI Systems: From Principles to Practice}},
 url = {https://iphome.hhi.de/samek/pdf/BerAudit22.pdf}
}

@misc{Bharadhwaj.2021,
 author = {Bharadhwaj, Homanga and Huang, De-An and Xiao, Chaowei and Anandkumar, Anima and Garg, Animesh},
 year = {2021},
 title = {{Auditing AI models for Verified Deployment under Semantic Specifications}},
 doi = {10.48550/arXiv.2109.12456}
}

@inproceedings{Bose.2019,
 author = {Bose, R P Jagadeesh C and Singi, Kapil and Kaulgud, Vikrant and Phokela, Kanchanjot K and Podder, Sanjay},
 title = {{Framework for Trustworthy Software Development}},
 pages = {45--48},
 publisher = {IEEE},
 booktitle = {{2019 34th IEEE/ACM International Conference on Automated Software Engineering Workshop (ASEW)}},
 year = {2019},
 address = {San Diego, CA},
 doi = {10.1109/ASEW.2019.00027}
}

@misc{Brundage.2020,
 author = {Brundage, Miles and Avin, Shahar and Wang, Jasmine and Belfield, Haydn and Krueger, Gretchen and Hadfield, Gillian and Khlaaf, Heidy and Yang, Jingying and Toner, Helen and Fong, Ruth and Maharaj, Tegan and Koh, Pang Wei and Hooker, Sara and Leung, Jade and Trask, Andrew and Bluemke, Emma and Lebensold, Jonathan and O'Keefe, Cullen and Koren, Mark and Ryffel, Th{\'e}o and Rubinovitz, J. B. and Besiroglu, Tamay and Carugati, Federica and Clark, Jack and Eckersley, Peter and de Haas, Sarah and Johnson, Maritza and Laurie, Ben and Ingerman, Alex and Krawczuk, Igor and Askell, Amanda and Cammarota, Rosario and Lohn, Andrew and Krueger, David and Stix, Charlotte and Henderson, Peter and Graham, Logan and Prunkl, Carina and Martin, Bianca and Seger, Elizabeth and Zilberman, Noa and h{\'E}igeartaigh, Se{\'a}n {\'O}. and Kroeger, Frens and Sastry, Girish and Kagan, Rebecca and Weller, Adrian and Tse, Brian and Barnes, Elizabeth and Dafoe, Allan and Scharre, Paul and Herbert-Voss, Ariel and Rasser, Martijn and Sodhani, Shagun and Flynn, Carrick and Gilbert, Thomas Krendl and Dyer, Lisa and Khan, Saif and Bengio, Yoshua and Anderljung, Markus},
 year = {2020},
 title = {{Toward Trustworthy AI Development: Mechanisms for Supporting Verifiable Claims}},
 doi = {10.48550/arXiv.2004.07213}
}

@misc{EC.2022,
 author = {{European Commission}},
 year = {2022},
 title = {{Draft standardisation request to the European Standardisation Organisations in support of safe and trustworthy artificial intelligence}}
}

@misc{EHLEG.2019,
 author = {{European High-Level Expert Group on AI}},
 year = {2019},
 title = {{Ethics guidelines for trustworthy AI | Shaping Europe's digital future}},
 url = {https://digital-strategy.ec.europa.eu/en/library/ethics-guidelines-trustworthy-ai},
 institution = {{European High-Level Expert Group on AI}}
}

@article{EitelPorter.2021,
 author = {Eitel-Porter, Ray},
 year = {2021},
 title = {{Beyond the promise: implementing ethical AI}},
 pages = {73--80},
 volume = {1},
 number = {1},
 issn = {2730-5953},
 journal = {{AI and Ethics}},
 doi = {10.1007/s43681-020-00011-6}
}

@book{ElEmam.2020,
 author = {{El Emam}, Khaled and Mosquera, Lucy and Hoptroff, Richard},
 year = {2020},
 title = {{Practical Synthetic Data Generation}},
 publisher = {{O'Reilly Media, Inc}},
 isbn = {9781492072744}
}

@article{Falco.2021,
 author = {Falco, Gregory and Shneiderman, Ben and Badger, Julia and Carrier, Ryan and Dahbura, Anton and Danks, David and Eling, Martin and Goodloe, Alwyn and Gupta, Jerry and Hart, Christopher and Jirotka, Marina and Johnson, Henric and LaPointe, Cara and Llorens, Ashley J. and Mackworth, Alan K. and Maple, Carsten and P{\'a}lsson, Sigur{\dh}ur Emil and Pasquale, Frank and Winfield, Alan and Yeong, Zee Kin},
 year = {2021},
 title = {{Governing AI safety through independent audits}},
 pages = {566--571},
 volume = {3},
 number = {7},
 issn = {2522-5839},
 journal = {{Nature Machine Intelligence}},
 doi = {10.1038/s42256-021-00370-7}
}

@misc{Felderer.2021,
 author = {Felderer, Michael and Ramler, Rudolf},
 year = {2021},
 title = {{Quality Assurance for AI-based Systems: Overview and Challenges}},
 doi = {10.48550/arXiv.2102.05351}
}

@inproceedings{GaldonClavell.2020,
 author = {{Galdon-Clavell}, Gemma and {Zamorano}, Mariano {M.} and Castillo, Carlos and Smith, Oliver and Matic, Aleksandar},
 title = {{Auditing Algorithms: On Lessons Learned and the Risks of Data Minimization}},
 pages = {265--271},
 publisher = {ACM},
 isbn = {978-1-4503-7110-0},
 booktitle = {{Proceedings of the AAAI/ACM Conference on AI, Ethics, and Society}},
 year = {2020},
 address = {New York},
 doi = {10.1145/3375627.3375852}
}

@inproceedings{Gardner.2019,
 author = {Gardner, Josh and Brooks, Christopher and Baker, Ryan},
 title = {{Evaluating the Fairness of Predictive Student Models Through Slicing Analysis}},
 pages = {225--234},
 publisher = {ACM},
 isbn = {978-1-4503-6256-6},
 series = {{LAK19}},
 booktitle = {{Proceedings of the 9th International Conference on Learning Analytics {\&} Knowledge}},
 year = {2019},
 address = {New York},
 doi = {10.1145/3303772.3303791}
}

@article{Garrison.1999,
 author = {Garrison, D. Randy and Anderson, Terry and Archer, Walter},
 year = {1999},
 title = {{Critical Inquiry in a Text-Based Environment: Computer Conferencing in Higher Education}},
 pages = {87--105},
 volume = {2},
 number = {2},
 issn = {1096-7516},
 journal = {{The Internet and Higher Education}},
 doi = {10.1016/S1096-7516(00)00016-6}
}

@article{Gebru.2021,
 author = {Gebru, Timnit and Morgenstern, Jamie and Vecchione, Briana and Vaughan, Jennifer Wortman and Wallach, Hanna and Iii, Hal Daum{\'e} and Crawford, Kate},
 year = {2021},
 title = {{Datasheets for datasets}},
 pages = {86--92},
 volume = {64},
 number = {12},
 issn = {0001-0782},
 journal = {{Communications of the ACM}},
 doi = {10.1145/3458723}
}

@misc{Green.2023,
 author = {Green, Marcus},
 year = {2023},
 title = {{Moodle 4.1 Database}},
 url = {https://www.examulator.com/er/4.1/index.html},
 urldate = {2023-01-24}
}

@article{Guidotti.2018,
 author = {Guidotti, Riccardo and Monreale, Anna and Ruggieri, Salvatore and Turini, Franco and Giannotti, Fosca and Pedreschi, Dino},
 year = {2018},
 title = {{A Survey of Methods for Explaining Black Box Models}},
 pages = {93:1--93:42},
 volume = {51},
 number = {5},
 issn = {0360-0300},
 journal = {{ACM Computing Surveys}},
 doi = {10.1145/3236009}
}

@misc{Holland.2018,
 author = {Holland, Sarah and Hosny, Ahmed and Newman, Sarah and Joseph, Joshua and Chmielinski, Kasia},
 year = {2018},
 title = {{The Dataset Nutrition Label: A Framework To Drive Higher Data Quality Standards}},
 doi = {10.48550/arXiv.1805.03677}
}

@inproceedings{Jacobs.2021,
 author = {Jacobs, Abigail Z. and Wallach, Hanna},
 title = {{Measurement and Fairness}},
 pages = {375--385},
 publisher = {ACM},
 isbn = {978-1-4503-8309-7},
 booktitle = {{Proceedings of the 2021 ACM Conference on Fairness, Accountability, and Transparency}},
 year = {2021},
 address = {Online},
 doi = {10.1145/3442188.3445901}
}

@inproceedings{Jentzsch.2019,
 author = {Jentzsch, Sophie F. and Hochgeschwender, Nico},
 title = {{Don't Forget Your Roots! Using Provenance Data for Transparent and Explainable Development of Machine Learning Models}},
 pages = {37--40},
 publisher = {IEEE},
 booktitle = {{2019 34th IEEE/ACM International Conference on Automated Software Engineering Workshop (ASEW)}},
 year = {2019},
 address = {San Diego},
 doi = {10.1109/ASEW.2019.00025}
}

@article{Kale.2022,
 author = {Kale, Amruta and Nguyen, Tin and {Harris, Frederick C., Jr.} and Li, Chenhao and Zhang, Jiyin and Ma, Xiaogang},
 year = {2022},
 title = {{Provenance documentation to enable explainable and trustworthy AI: A literature review}},
 pages = {1--41},
 issn = {2641-435X},
 journal = {{Data Intelligence}},
 doi = {10.1162/dint_a_00119}
}

@article{Kochling.2020,
 author = {K{\"o}chling, Alina and Wehner, Marius Claus},
 year = {2020},
 title = {{Discriminated by an algorithm: a systematic review of discrimination and fairness by algorithmic decision-making in the context of HR recruitment and HR development}},
 pages = {795--848},
 volume = {13},
 number = {3},
 issn = {2198-2627},
 journal = {{Business Research}},
 doi = {10.1007/s40685-020-00134-w}
}

@misc{Kreuzberger.2022,
 author = {Kreuzberger, Dominik and K{\"u}hl, Niklas and Hirschl, Sebastian},
 year = {2022},
 title = {{Machine Learning Operations (MLOps): Overview, Definition, and Architecture}},
 doi = {10.48550/arXiv.2205.02302}
}

@inproceedings{Mitchell.2019,
 author = {Mitchell, Margaret and Wu, Simone and Zaldivar, Andrew and Barnes, Parker and Vasserman, Lucy and Hutchinson, Ben and Spitzer, Elena and Raji, Inioluwa Deborah and Gebru, Timnit},
 title = {{Model Cards for Model Reporting}},
 pages = {220--229},
 publisher = {{Association for Computing Machinery}},
 isbn = {978-1-4503-6125-5},
 series = {{FAT* '19}},
 booktitle = {{Proceedings of the Conference on Fairness, Accountability, and Transparency}},
 year = {2019},
 address = {New York},
 doi = {10.1145/3287560.3287596}
}

@article{Mokander.2021,
 author = {M{\"o}kander, Jakob and Floridi, Luciano},
 year = {2021},
 title = {{Ethics-Based Auditing to Develop Trustworthy AI}},
 pages = {323--327},
 volume = {31},
 number = {2},
 issn = {0924-6495},
 journal = {{Minds and Machines}},
 doi = {10.1007/s11023-021-09557-8}
}

@article{Mokander.2022,
 author = {M{\"o}kander, Jakob and Axente, Maria and Casolari, Federico and Floridi, Luciano},
 year = {2022},
 title = {{Conformity Assessments and Post-market Monitoring: A Guide to the Role of Auditing in the Proposed European AI Regulation}},
 pages = {241--268},
 pagination = {page},
 volume = {32},
 issn = {0924-6495},
 journal = {{Minds \& Machines}},
 language = {eng},
 %shorthand = {Conformity Assessments and Post-market Monitoring: A Guide to the Role of Auditing in the Proposed European AI Regulation},
 doi = {10.1007/s11023-021-09577-4},
 number = {2},
 eprint = {34754142}
}

@inproceedings{Monllao.2018,
 author = {{Monlla{\'o} Oliv{\'e}}, David and {Du Huynh}, Q. and Reynolds, Mark and Dougiamas, Martin and Wiese, Damyon},
 title = {{A supervised learning framework for learning management systems}},
 pages = {1--8},
 publisher = {{Association for Computing Machinery}},
 isbn = {978-1-4503-6536-9},
 series = {{DATA '18}},
 booktitle = {{Proceedings of the First International Conference on Data Science, E-learning and Information Systems}},
 year = {2018},
 address = {New York},
 doi = {10.1145/3279996.3280014}
}

@misc{Moodle.Documentation,
 author = {Moodle},
 year = {2023},
 title = {{Documentation}},
 url = {https://docs.moodle.org/403/en/},
 urldate = {2023-12-20}
}

@misc{Moodle.AIprinciples,
 author = {Moodle},
 year = {2023},
 title = {{Moodle and our AI principles}},
 url = {https://moodle.com/moodle-and-our-ai-principles/},
 urldate = {2023-09-27}
}

@misc{Moodle.AnalyticsAPI,
 author = {Moodle},
 year = {2022},
 title = {{Analytics API}},
 url = {https://moodledev.io/docs/apis/subsystems/analytics/},
 urldate = {2023-01-13}
}

@article{Naja.2022,
 author = {Naja, Iman and Markovic, Milan and Edwards, Peter and Pang, Wei and Cottrill, Caitlin and Williams, Rebecca},
 year = {2022},
 title = {{Using Knowledge Graphs to Unlock Practical Collection, Integration, and Audit of AI Accountability Information}},
 pages = {74383--74411},
 volume = {10},
 issn = {2169-3536},
 journal = {{IEEE Access}},
 doi = {10.1109/ACCESS.2022.3188967}
}

@incollection{Prinsloo.2017,
 author = {Prinsloo, Paul and Slade, Sharon},
 title = {{Ethics and Learning Analytics: Charting the (Un)Charted}},
 pages = {49--57},
 publisher = {{Society for Learning Analytics Research (SoLAR)}},
 isbn = {978-0-9952408-0-3},
 editor = {Lang, Charles and Siemens, George and Wise, Alyssa and Gasevic, Dragan and {University of Edinburgh}, U. K.},
 booktitle = {{Handbook of Learning Analytics}},
 year = {2017},
 doi = {10.18608/hla17.004}
}

@inproceedings{Raghavan.2020,
 author = {Raghavan, Manish and Barocas, Solon and Kleinberg, Jon and Levy, Karen},
 title = {{Mitigating bias in algorithmic hiring: evaluating claims and practices}},
 pages = {469--481},
 publisher = {ACM},
 isbn = {978-1-4503-6936-7},
 series = {{FAT* '20}},
 booktitle = {{Proceedings of the 2020 Conference on Fairness, Accountability, and Transparency}},
 year = {2020},
 doi = {10.1145/3351095.3372828}
}

@inproceedings{Raji.2020,
 author = {Raji, Inioluwa Deborah and Smart, Andrew and White, Rebecca N. and Mitchell, Margaret and Gebru, Timnit and Hutchinson, Ben and Smith-Loud, Jamila and Theron, Daniel and Barnes, Parker},
 title = {{Closing the AI accountability gap: defining an end-to-end framework for internal algorithmic auditing}},
 pages = {33--44},
 publisher = {ACM},
 isbn = {978-1-4503-6936-7},
 series = {{FAT* '20}},
 booktitle = {{Proceedings of the 2020 Conference on Fairness, Accountability, and Transparency}},
 year = {2020},
 doi = {10.1145/3351095.3372873}
}

@inproceedings{Rzepka.2022,
 author = {Rzepka, Nathalie and Simbeck, Katharina and M{\"u}ller, Hans-Georg and Pinkwart, Niels},
 title = {{Fairness of In-session Dropout Prediction}},
 pages = {316--326},
 publisher = {{Scitepress}},
 isbn = {978-989-758-562-3},
 booktitle = {{Proceedings of the 14th International Conference on Computer Supported Education (CSEDU)}},
 year = {2022},
 doi = {10.5220/0010962100003182}
}

@article{Rzepka.2023,
 author = {Rzepka, Nathalie and Fernsel, Linda and M{\"u}ller, Hans-Georg and Simbeck, Katharina and Pinkwart, Niels},
 year = {2023},
 title = {{Unbias me! Mitigating Algorithmic Bias for Less-studied Demographic Groups in the Context of Language  Learning Technology}},
 pages = {1--23},
 volume = {6},
 number = {1},
 journal = {{Computer-Based Learning in Context}},
 doi = {10.5281/zenodo.7996194}
}

@article{SAI.2023,
 author = {Beckstrom, Jan Roar},
 year = {2021},
 title = {{Auditing machine learning algorithms. A white paper for public auditors}},
 pages = {40--41},
 volume = {48},
 number = {1},
 journal = {{International Journal of Government Auditing}}
}

@article{Shneiderman.2020,
 author = {Shneiderman, Ben},
 year = {2020},
 title = {{Human-Centered Artificial Intelligence: Three Fresh Ideas}},
 pages = {109--124},
 issn = {19443900},
 journal = {{AIS Transactions on Human-Computer Interaction}},
 doi = {10.17705/1thci.00131}
}

@article{Simbeck.2023,
 author = {Simbeck, Katharina},
 year = {2024},
 title = {{They shall be fair, transparent, and robust: auditing learning analytics systems}},
 issn = {2730-5953},
 journal = {{AI and Ethics}},
 volume = {4},
 doi = {10.1007/s43681-023-00292-7}
}

@book{Slade.2019,
 author = {Slade, Sharon and Tait, Alan},
 year = {2019},
 title = {{Global guidelines: Ethics in Learning Analytics}},
 publisher = {{ICDE}}
}

@misc{Soler.2023,
 author = {{Soler Garrido}, Josep and Tolan, Songul and {Hupont Torres}, Isabelle and {Fernandez Llorca}, David and Charisi, Vasiliki and {Gomez Gutierrez}, Emilia and Junklewitz, Henrik and Hamon, Ronan and {Fano Yela}, Delia and Panigutti, Cecilia},
 year = {2023},
 title = {{AI Watch: Artificial Intelligence Standardisation Landscape Update: JRC Research Reports}},
 doi = {10.2760/131984}
}

@inproceedings{Springer.2019,
 author = {Springer, Aaron and Whittaker, Steve},
 title = {{Making Transparency Clear: The Dual Importance of Explainability and Auditability}},
 pages = {4},
 publisher = {ACM},
 booktitle = {{Joint Proceedings of the ACM IUI 2019 Workshops}},
 year = {2019},
 address = {Los Angeles}
}

@article{Stoel.2012,
 author = {Stoel, Dale and Havelka, Douglas and Merhout, Jeffrey W.},
 year = {2012},
 title = {{An analysis of attributes that impact information technology audit quality: A study of IT and financial audit practitioners}},
 pages = {60--79},
 volume = {13},
 number = {1},
 issn = {14670895},
 journal = {{International Journal of Accounting Information Systems}},
 doi = {10.1016/j.accinf.2011.11.001}
}

@inproceedings{Suresh.2021,
 author = {Suresh, Harini and Guttag, John},
 title = {{A Framework for Understanding Sources of Harm throughout the Machine Learning Life Cycle}},
 pages = {1--9},
 publisher = {{Association for Computing Machinery}},
 isbn = {978-1-4503-8553-4},
 series = {{EAAMO '21}},
 booktitle = {{Equity and Access in Algorithms, Mechanisms, and Optimization}},
 year = {2021},
 address = {New York},
 doi = {10.1145/3465416.3483305}
}

@inproceedings{Tagharobi.2022,
 author = {Tagharobi, Hassan and Simbeck, Katharina},
 title = {{Introducing a Framework for Code based Fairness Audits of Learning Analytics Systems on the Example of Moodle Learning Analytics}},
 pages = {45--55},
 volume = {2},
 publisher = {{Scitepress}},
 isbn = {978-989-758-562-3},
 booktitle = {{Proceedings of the 14th International Conference on Computer Supported Education (CSEDU)}},
 year = {2022},
 doi = {10.5220/0010998900003182}
}

@article{Tao.2019,
 author = {Tao, Chuanqi and Gao, Jerry and Wang, Tiexin},
 year = {2019},
 title = {{Testing and Quality Validation for AI Software--Perspectives, Issues, and Practices}},
 pages = {120164--120175},
 volume = {7},
 issn = {2169-3536},
 journal = {{IEEE Access}},
 doi = {10.1109/ACCESS.2019.2937107}
}

@misc{Toreini.2022,
 author = {Toreini, Ehsan and Aitken, Mhairi and Coopamootoo, Kovila P. L. and Elliott, Karen and Zelaya, Vladimiro Gonzalez and Missier, Paolo and Ng, Magdalene and {van Moorsel}, Aad},
 year = {2022},
 title = {{Technologies for Trustworthy Machine Learning: A Survey in a Socio-Technical Context}},
 doi = {10.48550/arXiv.2007.08911 }
}

@inproceedings{Weigand.2013,
 author = {Weigand, Hans and Johannesson, Paul and Andersson, Birger and Bergholtz, Maria},
 title = {{Conceptualizing Auditability}},
 pages = {8},
 publisher = {CEUR},
 editor = {Deneck{\`e}re, R{\'e}becca and Proper, Henderik A.},
 booktitle = {{Proceedings of the CAiSE'13 Forum at the 25th International Conference on Advanced  Information Systems Engineering (CAiSE)}},
 year = {2013},
 address = {Valencia, Spain}
}

@misc{WH.2022,
 author = {{White House}},
 year = {2022},
 title = {{Blueprint for an AI Bill of Rights: Making Automated Systems Work for the American People}},
 url = {https://www.whitehouse.gov/ostp/ai-bill-of-rights/},
 urldate = {2023-12-20}
}

@article{Williams.2022,
 author = {Williams, Rebecca and Cloete, Richard and Cobbe, Jennifer and Cottrill, Caitlin and Edwards, Peter and Markovic, Milan and Naja, Iman and Ryan, Frances and Singh, Jatinder and Pang, Wei},
 year = {2022},
 title = {{From transparency to accountability of intelligent systems: Moving beyond aspirations}},
 volume = {4},
 number = {2022},
 journal = {{Data {\&} Policy}},
 doi = {10.1017/dap.2021.37}
}

@book{Wolnizer.2006,
 author = {Wolnizer, Peter W.},
 year = {2006},
 title = {{Auditing as Independent Authentication}},
 address = {Sydnesy},
 publisher = {{Sydney University Press}},
 isbn = {978-1-920898-34-2}
}

@article{zhidkikh2024reproducing,
  title={Reproducing Predictive Learning Analytics in CS1: Toward Generalizable and Explainable Models for Enhancing Student Retention.},
  author={Zhidkikh, Denis and Heilala, Ville and Van Petegem, Charlotte and Dawyndt, Peter and Jarvinen, Miitta and Viitanen, Sami and De Wever, Bram and Mesuere, Bart and Lappalainen, Vesa and Kettunen, Lauri and others},
  journal={Journal of Learning Analytics},
  volume={11},
  number={1},
  pages={132--150},
  year={2024},
  publisher={ERIC}
}

@article{van2023pass,
  title={Pass/fail prediction in programming courses},
  author={Van Petegem, Charlotte and Deconinck, Louise and Mourisse, Dieter and Maertens, Rien and Strijbol, Niko and Dhoedt, Bart and De Wever, Bram and Dawyndt, Peter and Mesuere, Bart},
  journal={Journal of Educational Computing Research},
  volume={61},
  number={1},
  pages={68--95},
  year={2023},
  doi={10.1177/07356331221085595  }
}

@article{Alfredo.2024,
 author = {Alfredo, Riordan and Echeverria, Vanessa and Jin, Yueqiao and Yan, Lixiang and Swiecki, Zachari and Ga{\v{s}}evi{\'c}, Dragan and Martinez-Maldonado, Roberto},
 year = {2024},
 title = {{Human-centred Learning Analytics and AI in Education: A systematic Literature Review}},
 pages = {100215},
 volume = {6},
 issn = {2666920X},
 journal = {{Computers and Education: Artificial Intelligence}},
 doi = {10.1016/j.caeai.2024.100215}
}

@article{Baek.2023,
 author = {Baek, Clare and Doleck, Tenzin},
 year = {2023},
 title = {{Educational Data Mining versus Learning Analytics: A Review of Publications From 2015 to 2019}},
 pages = {3828--3850},
 volume = {31},
 number = {6},
 issn = {1049-4820},
 journal = {{Interactive Learning Environments}},
 doi = {10.1080/10494820.2021.1943689}
}

@article{Drugova.2024,
 author = {Drugova, Elena and Zhuravleva, Irina and Zakharova, Ulyana and Latipov, Adel},
 year = {2024},
 title = {{Learning Analytics driven Improvements in Learning Design in higher Education: A systematic Literature Review}},
 pages = {510--524},
 volume = {40},
 number = {2},
 journal = {{Journal of Computer Assisted Learning}},
 doi = {10.1111/jcal.12894}
}

@article{Elmoazen.2023,
 author = {Elmoazen, Ramy and Saqr, Mohammed and Khalil, Mohammad and Wasson, Barbara},
 year = {2023},
 title = {{Learning Analytics in virtual Laboratories: A systematic Literature Review of empirical Research}},
 volume = {10},
 number = {1},
 journal = {{Smart Learning Environments}},
 doi = {10.1186/s40561-023-00244-y}
}

@article{Khalil.2023,
 author = {Khalil, Mohammad and Slade, Sharon and Prinsloo, Paul},
 year = {2023},
 title = {{Learning Analytics in Support of Inclusiveness and disabled Students: A systematic Review}},
 pages = {202--219},
 issn = {1042-1726},
 journal = {{Journal of Computing in Higher Education}},
 doi = {10.1007/s12528-023-09363-4}
}

@article{Minkkinen.2022,
 author = {Minkkinen, Matti and Laine, Joakim and M{\"a}ntym{\"a}ki, Matti},
 year = {2022},
 title = {{Continuous Auditing of Artificial Intelligence: A Conceptualization and Assessment of Tools and Frameworks}},
 volume = {1},
 number = {3},
 journal = {{Digital Society}},
 doi = {10.1007/s44206-022-00022-2}
}

@article{Ouyang.2023,
 author = {Ouyang, Fan and Wu, Mian and Zheng, Luyi and Zhang, Liyin and Jiao, Pengcheng},
 year = {2023},
 title = {{Integration of artificial intelligence performance prediction and learning analytics to improve student learning in online engineering course}},
 pages = {4},
 volume = {20},
 number = {1},
 journal = {{International Journal of Educational Technology in Higher Education}},
 doi = {10.1186/s41239-022-00372-4 }
}

@article{Sghir.2022,
 author = {Sghir, Nabila and Adadi, Amina and Lahmer, Mohammed},
 year = {2022},
 title = {{Recent advances in Predictive Learning Analytics: A decade systematic review (2012-2022)}},
 pages = {1--35},
 issn = {1360-2357},
 journal = {{Education and Information Technologies}},
 doi = {10.1007/s10639-022-11536-0}
}

@article{Xiong.2024,
 author = {Xiong, Zhang and Li, Haoxuan and Liu, Zhuang and Chen, Zhuofan and Zhou, Hao and Rong, Wenge and Ouyang, Yuanxin},
 year = {2024},
 title = {{A Review of Data Mining in Personalized Education: Trends and Future Prospects Current}},
 volume = {1},
 number = {26-50},
 journal = {{Frontiers of Digital Education}},
 doi = {10.3868/s110-009-024-0004-9}
}

@article{Paolucci.2024,
 author = {Paolucci, Catherine and Vancini, Sam and {Bex Ii}, Richard T. and Cavanaugh, Catherine and Salama, Christine and de Araujo, Zandra},
 year = {2024},
 title = {{A review of learning analytics opportunities and challenges for K-12 education}},
 pages = {e25767},
 pagination = {page},
 volume = {10},
 issn = {2405-8440},
 journal = {{Heliyon}},
 language = {eng},
 %shorthand = {A review of learning analytics opportunities and challenges for K-12 education},
 doi = {10.1016/j.heliyon.2024.e25767 },
 number = {4},
 abstract = {Use of learning analytics to improve educational outcomes is a growing area of research. While learning analytics research has been more prevalent in higher education contexts, this study presents findings from a qualitative metasynthesis of 47 research publications related to opportunities and challenges relevant to learning analytics design, implementation, and research at the PK-12 level. Our findings indicate that, while many see the educational benefits of learning analytics (e.g., more equitable instruction, individualized learning, enhanced assessment for learning); others remain unconvinced by a lack of evidence of improved outcomes and concerned about persistent challenges and potentially harmful impacts (e.g., infringement on users' privacy, misuse or misinterpretation of data by educators). We conclude by considering implications for mathematics education stemming from our analysis of learning analytics and posing questions to shape future research and key developments in mathematics education as learning analytics become more prevalent.},
 file = {Paolucci, Vancini et al. 2024 - A review of learning analytics:Attachments/Paolucci, Vancini et al. 2024 - A review of learning analytics.pdf:application/pdf},
 eprint = {38390101}
}

@incollection{Fernsel.2024,
 author = {Fernsel, Linda and Kalff, Yannick and Simbeck, Katharina},
 title = {{Where is the evidence?}},
 keywords = {Journal},
 pages = {262--269},
 bookpagination = {page},
 volume = {2},
 isbn = {978-989-758-697-2 },
 editor = {Poquet, Oleksandra and Ortega-Arranz, ALejandro and Viberg, Olga and Chounta, Irene-Angelica and McLaren, Bruce M. and Jovanovic, Jelena},
 booktitle = {{Proceedings of the 16th International Conference on Computer Supported Education (CSEDU 2024)}},
 year = {2024},
 abstract = {},
 doi = {10.5220/0012689800003693 },
 usera = {db},
 %shorthand = {Where is the evidence?},
 subtitle = {{A plugin for auditing Moodle's Learning Analytics}},
 location = {Angers, Frankreich},
 volumes = {2},
 file = {Fernsel, Kalff et al 2024 - Where is the evidence:Attachments/Fernsel, Kalff et al 2024 - Where is the evidence.pdf:application/pdf}
}

@article{Ayling.2022,
 author = {Ayling, Jacqui and Chapman, Adriane},
 year = {2022},
 title = {{Putting AI Ethics to Work}},
 pages = {405--429},
 pagination = {page},
 volume = {2},
 journal = {{AI and Ethics}},
 subtitle = {{Are the Tools fit for Purpose?}},
 %shorthand = {Putting AI Ethics to Work},
 doi = {10.1007/s43681-021-00084-x},
 number = {3},
 abstract = {Bias, unfairness and lack of transparency and accountability in Artificial Intelligence (AI) systems, and the potential for the misuse of predictive models for decision-making have raised concerns about the ethical impact and unintended consequences of new technologies for society across every sector where data-driven innovation is taking place. This paper reviews the landscape of suggested ethical frameworks with a focus on those which go beyond high-level statements of principles and offer practical tools for application of these principles in the production and deployment of systems. This work provides an assessment of these practical frameworks with the lens of known best practices for impact assessment and audit of technology. We review other historical uses of risk assessments and audits and create a typology that allows us to compare current AI ethics tools to Best Practices found in previous methodologies from technology, environment, privacy, finance and engineering. We analyse current AI ethics tools and their support for diverse stakeholders and components of the AI development and deployment lifecycle as well as the types of tools used to facilitate use. From this, we identify gaps in current AI ethics tools in auditing and risk assessment that should be considered going forward.},
 file = {Ayling, Chapman 2022 - Putting AI Ethics to Work:Attachments/Ayling, Chapman 2022 - Putting AI Ethics to Work.pdf:application/pdf}
}

@article{Landers.2023,
 author = {Landers, Richard N. and Behrend, Tara S.},
 year = {2023},
 title = {{Auditing the AI Auditors. A Framework for Evaluating Fairness and Bias in high stakes AI predictive Models}},
 pages = {36--49},
 pagination = {page},
 volume = {78},
 journal = {{The American psychologist}},
 language = {eng},
 doi = {10.1037/amp0000972},
 number = {1},
 eprint = {35157476}
}

@article{Li.2024,
 author = {Li, Yueqi and Goel, Sanjay},
 year = {2024},
 title = {{Artificial Intelligence Auditability and Auditor Readiness for Auditing Artificial Intelligence Systems}},
 journal = {{SSRN Journal}},
 %shorthand = {Artificial Intelligence Auditability and Auditor Readiness for Auditing Artificial Intelligence Systems},
 doi = {10.2139/ssrn.4787236},
 abstract = {As the business community races to implement artificial intelligence (AI), there are several challenges that need to be addressed such as fairness and biases, transparency, denial of individual rights, and dilution of privacy. AI audits are expected to ensure that AI systems function lawfully, robustly, and follow ethical standards (e.g., fairness). While the auditability for financial audits and information system audits has been well addressed in the literature, auditability of AI systems has not been sufficiently addressed. AI auditability and auditors' competencies are crucial for ensuring AI audits are conducted with high quality. Research on the auditability of AI and the competencies of AI auditors is gravely lacking leaving risks in AI systems unmitigated. The primary reason is that the field is nascent and the rapid growth has left the audit profession struggling to catch up. Foundational work on establishing parameters for such research would help advance this research. In this paper, we explore AI auditability measures and competencies required for conducting AI audits. We conducted semi--structured interviews with 23 experienced AI professionals who have direct involvement or indirect exposure to AI audits. Based on our findings, we propose a framework of AI auditability and identify the competencies required to conduct AI audits. Our study serves as the first formal attempt to systematically identify and classify auditability measures and auditors' expertise demanded for AI audits based on practitioners' perspectives. Our findings contribute to the AI audit literature, inform AI developers about implementing auditability, guide the training of new AI auditors, and establish a foundation for further research in the field.},
 file = {Li, Goel 2024 - Artificial Intelligence Auditability and Auditor:Attachments/Li, Goel 2024 - Artificial Intelligence Auditability and Auditor.pdf:application/pdf}
}

@article{Minkkinen.2024,
 author = {Minkkinen, Matti and Niukkanen, Anniina and M{\"a}ntym{\"a}ki, Matti},
 year = {2024},
 title = {{What about investors? ESG analyses as tools for ethics-based AI auditing}},
 pages = {329--343},
 pagination = {page},
 volume = {39},
 journal = {{AI {\&} Society}},
 %shorthand = {What about investors? ESG analyses as tools for ethics-based AI auditing},
 doi = {10.1007/s00146-022-01415-0},
 number = {1},
 abstract = {},
 file = {Minkkinen, Niukkanen et al. 2022 - What about investors:Attachments/Minkkinen, Niukkanen et al. 2022 - What about investors.pdf:application/pdf}
}

@article{Alagic.2021,
 author = {Alagi{\'c}, Amra and Turulja, Lejla and Bajgori{\'c}, Nijaz},
 year = {2021},
 title = {{Identification of Information System Audit Quality Factors}},
 pages = {1--28},
 volume = {1},
 number = {2},
 journal = {{Journal of Forensic Accounting Profession}},
 doi = {10.2478/jfap-2021-0006},
}

@article{Knechel.2013,
 author = {Knechel, W. Robert and Krishnan, Gopal V. and Pevzner, Mikhail and Shefchik, Lori B. and Velury, Uma K.},
 year = {2013},
 title = {{Audit Quality: Insights from the Academic Literature}},
 pages = {385--421},
 volume = {32},
 number = {Supplement 1},
 issn = {0278-0380},
 journal = {{AUDITING: A Journal of Practice {\&} Theory}},
 doi = {10.2308/ajpt-50350}
}

@article{kitto2019practical,
  title={Practical ethics for building learning analytics},
  author={Kitto, Kirsty and Knight, Simon},
  journal={British Journal of Educational Technology},
  volume={50},
  number={6},
  pages={2855--2870},
  year={2019},
  publisher={Wiley Online Library}
}

@article{jones2019reconsidering,
  title={Reconsidering data in learning analytics: Opportunities for critical research using a documentation studies framework},
  author={Jones, Kyle ML and McCoy, Chase},
  journal={Learning, Media and Technology},
  volume={44},
  number={1},
  pages={52--63},
  year={2019},
  doi={10.1080/17439884.2018.1556216}
}

@inproceedings{haim2023open,
  title={How to open science: A principle and reproducibility review of the learning analytics and knowledge conference},
  author={Haim, Aaron and Shaw, Stacy and Heffernan, Neil},
  booktitle={LAK23: 13th International Learning Analytics and Knowledge Conference},
  pages={156--164},
  year={2023}
}

@inproceedings{ahn2021co,
  title={Co-designing for privacy, transparency, and trust in K-12 learning analytics},
  author={Ahn, June and Campos, Fabio and Nguyen, Ha and Hays, Maria and Morrison, Jan},
  booktitle={LAK21: 11th international learning analytics and knowledge conference},
  pages={55--65},
  year={2021}
}

@inproceedings{veljanova2023operationalising,
  title={Operationalising Transparency as an Integral Value of Learning Analytics Systems--From Ethical and Data Protection to Technical Design Requirements},
  author={Veljanova, Hristina and Barreiros, Carla and Gosch, Nicole and Staudegger, Elisabeth and Ebner, Martin and Lindstaedt, Stefanie},
  booktitle={International Conference on Human-Computer Interaction},
  pages={546--562},
  year={2023},
  organization={Springer}
}
\end{document}